# Anomalous Phonon-mode Dependence in Polarized Raman Spectroscopy of Topological Weyl Semimetal TaP

*Kunyan Zhang[†], Xiaoqi Pang[‡], Tong Wang[‡], Fei Han[§], Shun-Li Shang[∥], Nguyen T. Hung[⊥],*

*Ahmad R. T. Nugraha[‡,#], Zi-Kui Liu[∥], Mingda Li[§,*], Riichiro Saito[‡,*] and Shengxi Huang[†,*]*

[†]Department of Electrical Engineering, The Pennsylvania State University, University Park, PA 16802, United States

[‡]Department of Physics, Tohoku University, Sendai 980-8578, Japan

[§]Department of Nuclear Science and Engineering, Massachusetts Institute of Technology, Cambridge, MA 02139, United States

[∥]Department of Materials Science and Engineering, The Pennsylvania State University, University Park, PA 16802, United States

[⊥]Frontier Research Institute for Interdisciplinary Sciences, Tohoku University, Sendai 980-8578, Japan

[#]Research Center for Physics, Indonesian Institute of Sciences, Tangerang Selatan 15314, Indonesia

ABSTRACT

Topological Weyl semimetals (WSMs) have attracted widespread interests due to the chiral Weyl fermions and surface Fermi arcs that enable unique optical and transport phenomena. In this work, we present angle-resolved Raman spectroscopy of TaP, a prototypical noncentrosymmetric WSM, for five excitation wavelengths ranging from 364 to 785 nm. The Raman active modes, $A_1$, $B_1^1$, and $B_1^2$ modes, exhibit two main unique features beyond the conventional Raman theory. First, the relative intensities of Raman active modes change as a function of the excitation wavelength. Second, angle-resolved polarized Raman spectra show systematic deviation from the Raman tensor theory. In particular, the $B_1^1$ mode is absent for 633 nm excitation, whereas the $B_1^2$ mode shows an unusual two-fold symmetry instead of a four-fold symmetry for 488, 532, and 633 nm excitations. These unconventional phenomena are attributed to the interference effect in the Raman process owing to the existence of multiple carrier pockets with almost the same energy but different symmetries.

# I. INTRODUCTION

Topological Weyl semimetal (WSM) is a novel phase of matter that offers the realization of the Weyl fermion, i.e. a massless solution to the Dirac equation with definite chirality [1-5]. In the electronic bandstructure of WSMs, linear dispersive low-lying electronic excitations of Weyl fermions are formed as a natural consequence of the nontrivial topology of the band touching points termed Weyl nodes. The Weyl nodes appear in pairs with opposite chirality, and the momentum-space projections of the paired Weyl nodes on the surface are connected by the surface Fermi arc, where the density of states are open segments instead of a closed contour as conventional Fermi liquids. Moreover, in a WSM, a prominent phenomenon of chiral anomaly can be demonstrated, where the quantum-mechanical breaking of the classical chiral symmetry [6,7] leads to numerous novel observable phenomena, such as negative magnetoresistance [8-10], nonlocal transport [11], unconventional plasmon mode [12], phonon anomaly [13], second-harmonic generation [14], and circular photogalvanic effect [15,16]. Such exotic optical and electronic properties suggest that WSMs could be a promising platform for next-generation electronics and optoelectronics [17-21].

The isolation of a single Weyl node in WSMs demands the breaking of either inversion symmetry or time-reversal symmetry or both. Examples of WSMs that break inversion symmetry are the prototypical TaAs family (including TaAs, TaP, NbAs, and NbP) [4], $TlBiSe_2$ [22], $TaIrTe_4$ [23], etc. Among the TaAs family, TaP has the cleanest carrier pockets near the Fermi energy dominated by the Weyl fermions, that in return the topological states in TaP are easier to probe [24]. Extensive efforts have been made in identifying the topological states in TaP and unveiling its relevance to the electronic, magnetic and optical properties. Along with the angle-resolved photoelectron spectroscopy [25] and negative magnetoresistance [26], a strong intrinsic spin Hall effect in TaP has been predicted by *ab initio* calculations [27] and confirmed in experiments [28]. Recently, the giant thermoelectric response [29] and possible Kohn anomaly [30] at Weyl nodes have also been reported in TaP, which are intrinsically related to the topological signature of WSMs. Raman spectroscopy, though being underestimated in the field of topological materials, can offer unique insights from an electron-phonon interaction perspective with a phonon mode resolution, thus adding valuable insights to the electrical transport properties. The degenerate conduction bands in TaP would modify the resonant Raman spectra by the interference effect of Raman scattering process. Besides, we also expect Kohn anomaly effect in TaP because of the semimetallic energy dispersion similar to graphene [31].

Raman spectroscopy studies the light-matter interactions including electron-photon, electron-phonon, and electron-electron interactions [32,33], and thus serves as a probe for lattice dynamics. Liu, *et al*. [34,35] reported several Raman peaks in TaAs that belong to two-phonon excitation, which happens not only at the $\Gamma$ point but also at other k-points in the Brillouin zone. In addition, Raman spectroscopy can provide rich information on the crystal symmetry through the phonon symmetry and even monitor the phase transition due to structural change nondestructively. Liu *et al.* reported angle-resolved polarized Raman spectroscopy of WSM TaIrTe$_4$ in which a strong in-plane optical anisotropy is observed [36]. Another example is orthorhombic MoTe$_2$, a type-II WSM [37] exhibiting colossal magnetoresistance [38] and pressure-enhanced superconductivity [39]. Intriguingly, the topological states of MoTe$_2$ appear only at low temperatures (< 200 K) when it is isostructural with the noncentrosymmetric T$_d$ phase of WTe$_2$ [40-42]. The phase transition of MoTe$_2$ from topologically trivial 1T' phase to non-trivial T$_d$ phase has been witnessed by the emergence of new modes in the Raman spectra [43-46]. Thus, Raman spectroscopy of WSMs gives fundamental information on not only the phonon modes but also phase transitions and provides a useful gauge for the broken inversion symmetry.

In this work, we demonstrate how the polarized Raman spectra of TaP evolve with various excitation wavelengths which lead to different resonant scattering configurations. Five excitation wavelengths ranging from ultraviolet to visible spectra are employed: 364, 488, 532, 633, and 785 nm. Though little is shifted in the Raman frequency, Raman mode intensities of TaP show distinctive excitation wavelength dependence. In particular, the $B_1^2$ mode exhibits an evolution similar to the $A_1$ mode instead of the $B_1^1$ mode despite the same phonon symmetry. More intriguingly, the angle-resolved polarized Raman spectroscopy shows that the symmetry of the $B_1^2$ mode deviates from the predicted four-fold symmetry by the Raman tensor, whereas the $B_1^1$ mode is consistent with the Raman tensor theory. Besides the varied behaviors of the $B_1^1$ and $B_1^2$ modes, we found that, for 633 nm excitation, Raman intensity of the $B_1^1$ phonon mode at 190.5 cm$^{-1}$ in the experiments is significantly suppressed. The origin of the disappearance of Raman intensity can be explained by the Raman selection rule requiring that the initial, intermediate and final states involved in the resonance Raman process satisfy energy conditions. First-principles calculations based on density-functional theory (DFT), electron-phonon Wannier (EPW) package [47], and our own Raman intensity code [48] are employed for calculating resonant Raman spectra, which can well reproduce and qualitatively explain the anomalous aspects of the experimental observations. Our integrated experiment/theory studies on TaP unveil the optical spectroscopic responses and their relationship to the electronic structures of WSMs, which are essential for the endeavors of designing functional optoelectronics based on WSMs.

## II. RESULTS AND DISCUSSION

TaP belongs to body-centered tetragonal space group $I4_1md$ (No. 109) as shown in Fig. 1(a). TaP has 24 Weyl points with 8 of them on the $k_z = 0$ plane in the reciprocal space below Fermi level and 16 away from the $k_z = 0$ plane close to the Fermi level [25]. The reciprocal lattice of TaP is shown in Fig. 1(b), illustrating the high symmetry points of the Brillouin zone (Γ, Σ, S). In Figs. 1(c) and 1(d), we show the present electronic band structures of TaP without and with spin-orbit coupling (SOC), respectively, along high symmetry paths of the Brillouin zone (Γ-Σ-S, Fig. 1(b)), consistent with the previous calculations [25]. Our sample is synthesized using chemical vapor transport with high crystallinity (X-ray diffraction of the TaP crystal is shown in Fig. S1). In the Raman measurement, we adopt the $Z(XX)\overline{Z}$ configuration, which means that the incident and scattered light beams propagate along the Z and $\overline{Z}$ directions, respectively, and are selected to have the same polarization direction by a linear polarizer. Here, the experimental coordinates X and Y correspond to the $\textbf{\textit{a}}$-axis and $\textbf{\textit{b}}$-axis of the crystal, respectively, and the propagation directions Z and $\overline{Z}$ are along the $\textbf{\textit{c}}$-axis of the crystal. According to Raman tensor theory, the Raman intensity can be expressed as a function of the polarization angle $\theta$ in the XY plane, where $\theta$ is the angle of the incident laser polarization measured from the X-axis ($\textbf{\textit{a}}$-axis). The angular dependence of polarized Raman intensity is defined by

$$I = \left| \hat{e}_s{}^T \quad \widetilde{R} \quad \hat{e}_i \right|^2 \propto \left| (\cos\theta, \sin\theta, 0) \quad \widetilde{R} \quad \begin{pmatrix} \cos\theta \\ \sin\theta \\ 0 \end{pmatrix} \right|^2, \tag{1}$$

where $\widetilde{R}$ is the Raman tensor, $\hat{e}_i$ and $\hat{e}_s$ are the polarization vectors of the incident and scattered beams, respectively. Since there are two Ta and two P atoms in the primitive unit cell of TaP, it has twelve phonon modes in which four modes are doubly degenerate E modes. Vibrational modes of TaP consist of three acoustic phonon modes [$A_1 + E$] and nine optical phonon modes [$A_1 + 2B_1 + 3E$] where all the optical phonon modes are Raman active. The Raman intensity of the $A_1$ mode is calculated as follows:

$$I = \left| \hat{e}_s{}^T \quad \widetilde{R}(A_1) \quad \hat{e}_i \right|^2 = \left| (\cos\theta \quad \sin\theta \quad 0) \begin{pmatrix} a & 0 & 0 \\ 0 & a & 0 \\ 0 & 0 & c \end{pmatrix} \begin{pmatrix} \cos\theta \\ \sin\theta \\ 0 \end{pmatrix} \right|^2 = |a|^2, \tag{2}$$

meaning that the $A_1$ mode intensity does not have $\theta$ dependence. While the intensity of the $B_1$

mode varies with polarization angle as $|c \cos(2\theta)|^2$ because the Raman tensor of the $B_1$ mode takes a form of $\begin{pmatrix} c & 0 & 0 \\ 0 & -c & 0 \\ 0 & 0 & 0 \end{pmatrix}$. In order to distinguish two $B_1$ modes in TaP, we label them as $B_1^1$ and $B_1^2$, where the $B_1^2$ mode has a higher frequency than the lower $B_1^1$ mode. It is important to note that, the E mode is not observable for the $Z(XX)\overline{Z}$ configuration, since the Raman tensor of the E mode given by $\begin{pmatrix} 0 & 0 & e \\ 0 & 0 & 0 \\ e & 0 & 0 \end{pmatrix}$ and $\begin{pmatrix} 0 & 0 & 0 \\ 0 & 0 & e \\ 0 & e & 0 \end{pmatrix}$, respectively, renders zero Raman intensity.

In Table SI we summarize the polarization dependence of Raman intensity calculated by Raman tensor theory for the $Z(XX)\overline{Z}$ and other measurement configurations.

The present Raman measurement results are generally consistent with the analysis of Raman tensor theory, that is, only the $A_1$, $B_1^1$ and $B_1^2$ modes are observed at $\theta = 0°$ for $Z(XX)\overline{Z}$

configuration. In Fig. 2(a), we show Raman spectra measured with laser polarization along the *a*-

axis for five excitation wavelengths, 364, 488, 532, 633, and 785 nm. Three peaks are observed at

190.5, 373.9, and 411.7 cm$^{-1}$ and are assigned to the $B_1^1$, $A_1$ and $B_1^2$ modes, respectively. We also performed DFT calculations on the phonon modes and obtained the phonon frequencies at 179.5,

355.0, and 395.2 cm$^{-1}$ (Fig. 2(b)) in reasonably good agreement with the experimental values,

while the discrepancy arises mainly from (semi-)local DFT such as the exchange-correction (X-C)

functional [49]. For example, ancillary DFT calculations using the Vienna Ab initio Simulation

Package (VASP) and the X-C functional of generalized gradient approximations (GGA) with van der Waals correction yield three phonon frequencies of 194.5, 376.8, and 414.7 $cm^{-1}$, respectively. Raman frequencies of all the modes do not change with the laser wavelength, which is typical behavior of first-order Raman modes. However, the relative Raman intensities of the $B_1^1$, $A_1$ and $B_1^2$ modes conspicuously depend on the excitation wavelength. Though the intensity profiles for 364 and 532 nm excitations are similar to each other, particular modes have suppressed intensities for other excitation wavelengths, such as the $B_1^2$ mode for 488 nm and the $A_1$ mode for 785 nm (see insets of Fig. 2(a)). Unexpectedly, the $B_1^1$ mode is completely absent for 633 nm excitation, suggesting that the Raman mode intensities of TaP have distinctive excitation wavelength dependence.

We summarize the intensity evolution for each individual mode as a function of excitation wavelength in Fig. 2(c) for both the experiments (black line) and the DFT calculations (red line). All experimental values in Fig. 2(c) are normalized by the incident laser power with other measurement parameters set to be the same to yield an accurate comparison. The maximum intensities of the experimental $A_1$ and $B_1^2$ modes both occur for 633 nm which are 67 and 11 times higher than their intensity minima for 785 and 488 nm, respectively. However, the intensity of the $B_1^1$ mode is zero for 633 nm excitation, which is not consistent with the Raman tensor analysis in which we get a finite intensity $|c|^2$ at polarization angle $\theta = 0°$. Although the $B_1^1$ and $B_1^2$ modes represent the same phonon symmetry, they show almost opposite energy dependence in the experiments. In terms of the DFT calculations, Raman excitation profiles generally reproduce the experimental observations, especially for the $B_1^1$ mode. Considering the $B_1^1$ mode in Fig. 2(c), the Raman intensity first decreases to the minimum for 639 nm excitation and reaches to the maximum for 492 nm excitation. Aside from the numerical deviations for the $A_1$ and $B_1^2$ modes, the overall trends of the experimental results are similar to the DFT calculations: both intensities first increase then decrease to a plateau as the excitation wavelength decreases, which is opposite to the $B_1^1$ mode.

The resemblance between the $A_1$ and $B_1^2$ modes and the dissimilarity between the $B_1^1$ and $B_1^2$ modes indicate that the same phonon symmetry does not guarantee a similar Raman excitation profile, which is not consistent with the results in Table SI. Energy separations between electronic states involved in the interband transitions in resonant Raman scattering also contribute to the change of Raman intensity as a function of excitation wavelength. Resonant Raman scattering intensity can be explained by the third-order perturbation theory:

$$I(\omega_\nu) = \sum_i \left| \sum_{m,m'} \frac{\langle f|H_{op}|m'\rangle\langle m'|H_{ep}^\nu|m\rangle\langle m|H_{op}|i\rangle}{\left(E_L - E^{mi} - i\gamma\right)\left(E_L - E^{m'i} - \hbar\omega_\nu - i\gamma\right)} \right|^2, \tag{3}$$

where, $H_{ep}^\nu$ and $H_{op}$ are the electron-phonon interaction and the electron-photon interaction operators, respectively. In the denominator, $E_L$ is the laser energy, $E^{m(m')i} = E^{m(m')} - E^i$ is the energy difference between the initial state $i$ and the intermediate state $m(m')$, $\omega_\nu$ is the phonon mode frequency, and $\gamma = \frac{\hbar}{2\tau}$ is the broadening factor in the resonant Raman process related to the lifetime of the photo-excited carrier $\tau$. When $E^{mi}$ is close to the laser energy ($E_L = E^{mi} + i\gamma$), the interband transition is resonant with the incident photons and the Raman intensity is significantly enhanced. The resonance condition belongs to the incident resonance. In the case when $\gamma$ is large, the region of the resonant energy (or resonant window) increases, which is an important factor for analyzing the polarized Raman scattering of TaP as shown later. Another resonance condition is scattered resonance condition, $E_L - E^{m'i} - \hbar\omega_\nu - i\gamma = 0$, which relates to the phonon frequency $\omega_\nu$. In order to examine whether the incident resonance ($E_L - E^{mi} - i\gamma = 0$) plays an essential role in the Raman intensity, we performed spectroscopic ellipsometry measurement on TaP in the spectral range from 200 to 1000 nm as shown in Fig. 2(c) in which the imaginary part of permittivity positively correlates with the optical absorption. Since there is no peak in the absorption spectra, we do not expect different incident resonance for different excitation wavelengths. In addition to the resonant conditions, we need to consider the electron-phonon interaction $\langle m'|H_{ep}^\nu|m\rangle$ as a function of energy. Thus, the differences between the $B_1^1$ and $B_1^2$ modes suggest an important role of phonons in TaP and the effective coupling between electronic states and lattice vibrations (electron-phonon coupling). A similar analysis has been used to explain the excitation wavelength dependence of Raman modes in other quantum materials such as the giant Rashba material BiTeI [50], topological insulator $Bi_2Se_3$ [51], and Dirac semimetal ZrSiS [52]. Our results also imply that it is possible to control the electron-phonon coupling in the Raman scattering process of WSMs and to excite selectively the bands of interest, topologically nontrivial ones for example, by choosing an appropriate excitation wavelength.

Besides the interband transition resonance, the symmetry of lattice vibrations also affects the coupling of phonon modes to electronic states. To confirm the symmetry of lattice vibrations, we performed angle-resolved polarized Raman measurements and summarized the polarization dependence of the Raman intensities for every 15° rotation angle in Table I with the intensity magnitudes rescaled for better visual clarity. Here, $\theta = 0/180°$ and $\theta = 90/270°$ corresponds to the **a**-axis and **b**-axis of the crystal, respectively. Both the $B_1^1$ and $B_1^2$ mode intensities show anisotropy as expected and change in a periodic fashion, while the $A_1$ mode intensity does not strongly depend on $\theta$ for all excitations. The $B_1^1$ mode has precisely four-fold symmetry which can be expressed as $|c\cos(2\theta)|^2$ given by Raman tensor and the differences between the fitting values at $\theta = 0/180°$ and $\theta = 90/270°$ (i.e. the **a**-axis and **b**-axis) are less than 0.4%. However, the polarization dependence of the $B_1^2$ mode deviates from the four-fold symmetry: the $B_1^2$ mode intensity at $\theta = 0°/180°$ is evidently smaller than that at $\theta = 90°/270°$ for 488 and 532 nm excitations, while it is the opposite case for 633 nm. The intensity ratios of the $B_1^2$ mode between $\theta = 0°$ and $90°$ in the polar plot fittings are 1.00, 0.82, 0.72, 1.17, and 0.91 from 364 nm to 785 nm excitations. These intensity ratios show inconsistency with the Raman tensor analysis that is discussed in Table SI. Theoretically, TaP should give the same Raman intensity when the laser is polarized along the **a**- or **b**-axis. In the Raman study by Liu *et al*. on TaAs [34], another WSM with tetragonal space group, the measured Raman intensities of the $B_1^1$ and $B_1^2$ modes do not show observable disparity along the **a**- and **b**-axes for the same experimental configuration. Though they employed a single 514.5 nm laser as the excitation, different from our choice of wavelengths. More importantly, even though TaP and TaAs belong to the same space group and have the same crystal structure, they do not necessarily have the same Raman tensor because of different chemical compositions. Another intriguing observation is the missing $B_1^1$ mode for any polarization angle in the case of 633 nm excitation as presented in Figs. 3(a) and 3(b). In Fig. 3 we show 633 nm excitation color map and measured spectra, in which the $B_1^1$ mode is Raman inactive for all polarization angles. The polarized Raman spectra for the other four excitations are shown in Fig. S2.

To understand the unusual phonon symmetry of TaP, we performed first-principles DFT calculations for the polarization dependence of Raman intensity with different incident laser wavelengths. Here, we consider the same geometry of Raman scattering and we adopt five different excitation wavelengths that are near the experimental values including 366, 492, 536, 639, 794 nm. Overall, the DFT calculations (Table II) can reproduce the experiments (Table I) in terms of the relative magnitude and symmetry except for a few occasions. We have discussed the relationship between relative intensity and excitation wavelength with laser polarization along the *a*-axis and confirmed the accuracy of our program by the high resemblance (Fig. 2(c)). The angular dependence of the full width at half maximum (FWHM) is shown in the Supplemental Material Fig. S3. The following analysis will focus on the symmetry of the phonon modes. First, the angular dependence of the $A_1$ mode is highly isotropic for 492 and 536 nm excitation, while those for the other three laser excitations are slightly anisotropic. Similar features of the $A_1$ mode are observed in the measurement including the circular-shaped polar plots for 488 and 532 nm excitations and the elliptical-shaped polar plots for 364, 633, 785 nm excitations. For the $B_1^2$ modes, the DFT calculations also show a non-four-fold symmetry of the $B_1^2$ modes consistent with our Raman measurements. For example, the calculated $B_1^2$ mode for 536 nm excitation has the second maxima at the polarization angle $\theta=0°/180°$, same as the measured one for 532 nm excitation. However, there are some slight discrepancies between experiments and DFT calculations. For instance, the second maxima of the calculated $B_1^2$ mode for 639 nm excitation are rotated by $90°$ with respect to the measured result for 633 nm excitation. Moreover, the calculated relative Raman intensity of the $A_1$ mode is overestimated by one order of magnitude. Those discrepancies can be explained by adopting an adjustable parameter of $\gamma$ as explained below.

In Eq. (3), there are two energy denominators: $(E_L - E^{mi} - i\gamma)$ and $(E_L - E^{m'i} - \hbar\omega_\nu - i\gamma)$. When the resonance effect occurs, the Raman intensity strongly depends on the broadening factor $\gamma$. In other words, the intermediate states of the first-order Raman scattering processes will be changed when we choose different $\gamma$ values. It should be noted that the spectral width of Raman peaks (FWHM) does not come from the $\gamma$ factor of Eq. (3) but another $\Gamma$ factor for phonon frequency (not shown in Eq. (3)). The $\gamma$ factor in Eq. (3) affects only the resonant behavior as a function of laser excitation energy and thus the peak intensity, while the $\Gamma$ factor that modifies the phonon frequency affects the broadening of Raman peaks (FWHM) which is discussed in the Supplemental Material. Here, we focus on the $\gamma$ factor. For TaP, since the conduction bands are degenerate at some specific k-points in the Brillouin zone (Fig. 4(a)), two cases can happen. One is that the $\gamma$ value is much larger than the energy difference between two degenerate conduction bands, then both conduction bands contribute to the Raman intensity and the interference between two conduction bands also influences the Raman intensity. The other one is that the $\gamma$ value is much smaller than the energy difference between two degenerate conduction bands, only one conduction band mainly contributes to the Raman intensity and the interference between two conduction bands would not occur. In Fig. 4(b), we show the calculated polarization dependence for two different $\gamma$ values of 0.1 eV and 0.01 eV at the laser excitation of 2.52 eV (492 nm) and 1.94 eV (639 nm) for the $B_1^2$ Raman shift of 395.2 cm$^{-1}$. Both $\gamma$ values show that the four-fold symmetries are broken, and the maxima of Raman intensity shift is rotated from $90°$ to $0°$ for $\gamma = 0.1$ eV to 0.01 eV, respectively, while the Raman tensor theory shows a perfect four-fold symmetry. Since $\gamma = \frac{\hbar}{2\tau}$ is related to the inverse of the lifetime of photo-excited electron $\tau$, the $\gamma$ values depend on the laser energy. Therefore, the isotropic behavior of the $A_1$ mode and the near-four-fold symmetry of the $B_1$ mode only appear at specific laser wavelength excitation, which is consistent with our experiments (see Table I). It is important to mention that $\gamma$ can be phonon mode dependent since $\hbar\omega_\nu$ in the energy denominator ($E_L - E^{m'i} - \hbar\omega_\nu - i\gamma$) can have $\Delta\hbar\omega_\nu$ due to lifetime of phonons. In order to show the roles that quantum interference plays, we artificially remove the quantum interference effect in the calculation. That is, we calculate Raman intensity in which we intentionally select the cases of $m' = m$ in the summation of $m'$ in Eq.

(3) in the computational program. In this case, the deviation of the polarized Raman spectra does not happen for all phonon modes and for all laser energies (Fig. S4). Thus, the deviation occurs as the quantum interference effect when the contribution of $m' \neq m$ to Raman scattering amplitude is not negligible. As a special case, the quantum interference effect suppresses even the Raman intensity, which can be seen in the case of the $B_1^1$ mode for 633nm excitation.

## III. CONCLUSION

In conclusion, we measured the Raman responses of TaP using five laser excitation wavelengths (364, 488, 532, 633, and 785 nm) and studied its optical anisotropy by angle-resolved polarized Raman spectroscopy. From the measured results, we show the strong laser wavelength dependence and the deviation of the phonon symmetry of TaP, both of which cannot be fully explained by the conventional Raman tensor theory. The laser wavelength dependence does not solely rely on the symmetry of the phonon modes, as the $B_1^1$ and $B_1^2$ modes display different dependencies on laser wavelength. Polarized Raman spectroscopy shows that the $A_1$ mode is relatively isotropic and the $B_1^1$ and $B_1^2$ modes are two or four-fold symmetric. While the $B_1^1$ mode has a four-fold symmetry consistent with the Raman tensor theory, the $B_1^2$ mode shows a two-fold symmetry, especially for 488, 532, and 633 nm excitations. Most intriguingly, the $B_1^1$ mode is missing for 633 nm excitation as a result of quantum interference effect. Our refined DFT calculations provide a quantitative explanation for the abnormal phonon symmetry through the energy-dependent broadening factors and electron-phonon interactions. The integrated experimental measurements, calculations, and theoretical analyses provide useful insights into the electron-photon and electron-phonon interactions in TaP, as well as a practical method to understand WSMs, which is of great significance in exploring future applications in optoelectronics for this emerging class of quantum materials.

## METHOD

*Chemical Vapor Transport*: TaP single crystals with a typical size of 4×4×3 mm$^3$ are synthesized by chemical vapor transport method [29]. XRD experiment confirms that the crystals crystallize in the space group $I4_1md$ with lattice parameters $\boldsymbol{a}$ = 0.33281(2) nm, $\boldsymbol{c}$ = 1.13372(8) nm. Laue diffraction determines the orientations of the crystals.

*Raman spectroscopy*: Raman spectra are measured on a Horiba LabRam spectrometer. Five excitation wavelengths in the ultraviolet and visible regime are used in the Raman experiments: 364, 488, 532, 633, and 785 nm. Incident laser along the $\boldsymbol{c}$-axis is focused by 50× objective lens in a back-scattering configuration. Laser power ranges from 1 to 5 mW depending on the excitation wavelength to ensure sufficient signals without any visible oxidation on the sample surface. A polarization analyzer is coupled to the spectrometer to collect scattered light that is parallel-polarized to the incident light.

*DFT calculations*: The calculations of Raman intensity in TaP are based on DFT based first-principles calculations. First, we use the QuantumESPRESSO package [53] to obtain the electronic structure and phonon dispersion of TaP. Here, the potentials between ionic cores and valence electrons are modeled with norm-conserving pseudopotentials. The X-C energy of electrons is approximated within a local density approximation (LDA) [54]. The cutoff energy of 150 Ry is adopted to achieve the convergence of total energy. Using the calculated wavefunctions, we obtain the electron-photon matrix element by the home-made program [48]. Further, we use the electron-phonon Wannier package [47] to calculate the electron-phonon matrix elements for the phonon wavevector at the $\Gamma$ point in the Brillouin zone. In the calculations, the lattice parameters of TaP are optimized to $a$ = 0.33363 nm, $c$ = 1.14077 nm by the geometrical optimization, and is adopted as an adjustable parameter. In addition, ancillary DFT calculations are also performed to calculate phonon frequencies at $\Gamma$ points using the VASP code [55] in terms of a 72-atom supercell. Here the ion-electron interaction is described by the projector augmented wave method [56] and the X-C functional is described by the generalized gradient approximations (GGA) [57] incorporating with the van der Waals correction [58] (i.e., the GGA + D3 method).

**Supplemental Material**

See Supplemental Material at [URL will be inserted by publisher for [give brief description of material].

**Corresponding Author**

*E-mail: sjh5899@psu.edu (S. H.), rsaito@flex.phys.tohoku.ac.jp (R. S.), mingda@mit.edu (M.L.)

**Author Contributions**

The manuscript was written through the contributions of all authors. All authors have given approval to the final version of the manuscript.

ACKNOWLEDGMENT

Raman spectroscopy was performed at the Materials Characterization Lab at Pennsylvania State University. F.H. and M.L. acknowledge the support from U.S. DOE BES Award No. DE-SC0020148.

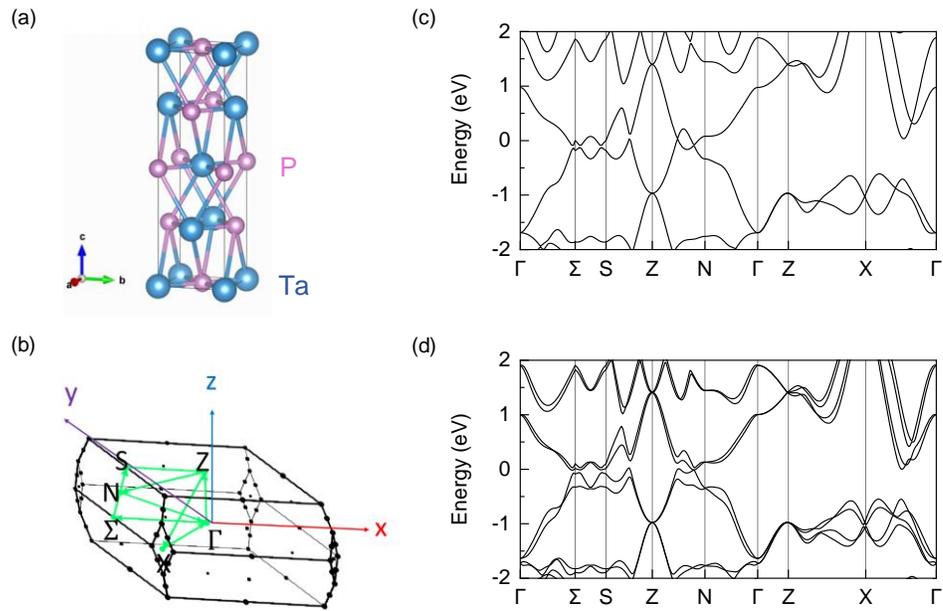

FIG. 1. Atomic structure and calculated band structure of TaP. (a) TaP atomic structure. (b) Schematics of the high symmetry lines in the Brillouin zone considered for the DFT calculations. (c) Calculated band structure without SOC and (d) with SOC using DFT.

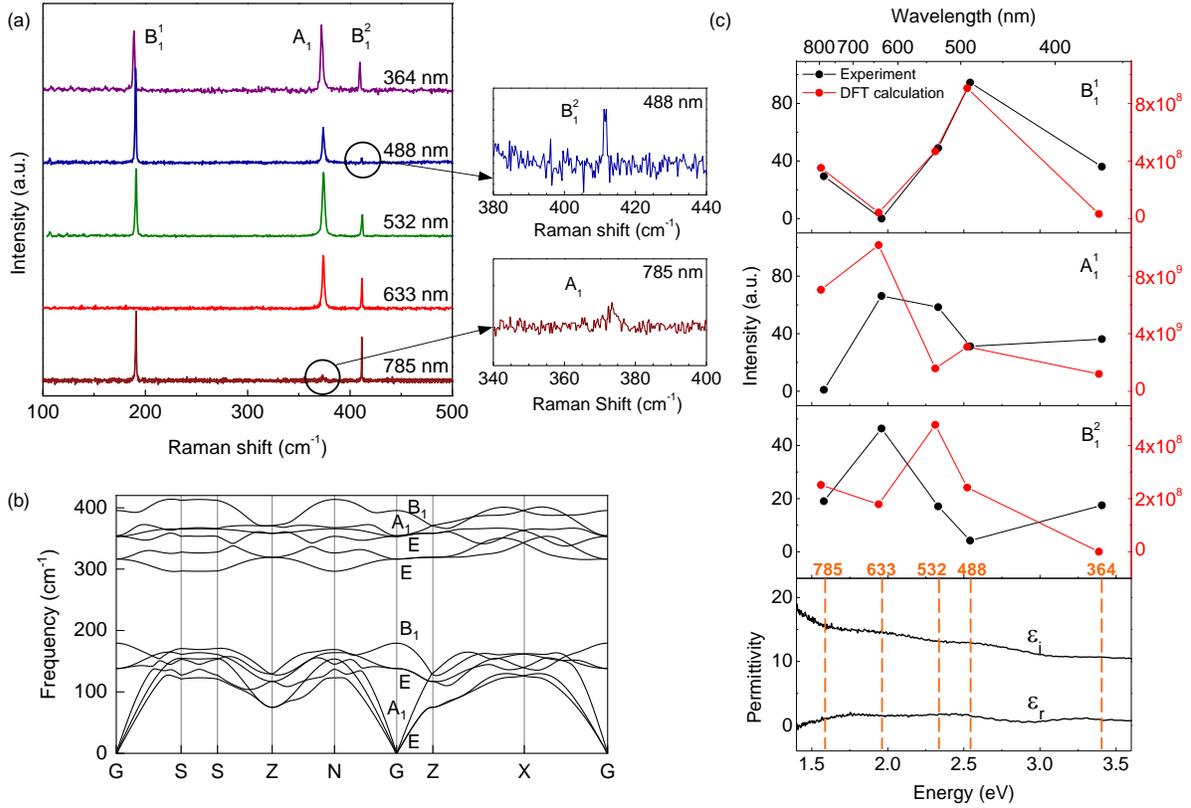

FIG. 2. Phonon modes and excitation wavelength dependence of Raman spectra of TaP. (a) Raman spectra for different excitation wavelengths. Laser polarization is along the *a*-axis and parallel-polarized scattered light is collected. (b) Calculated Raman modes and vibrational patterns. (c) Calculated and experimental excitation wavelength dependence of the Raman modes with laser polarization along the *a*-axis and the experimental permittivity of TaP.

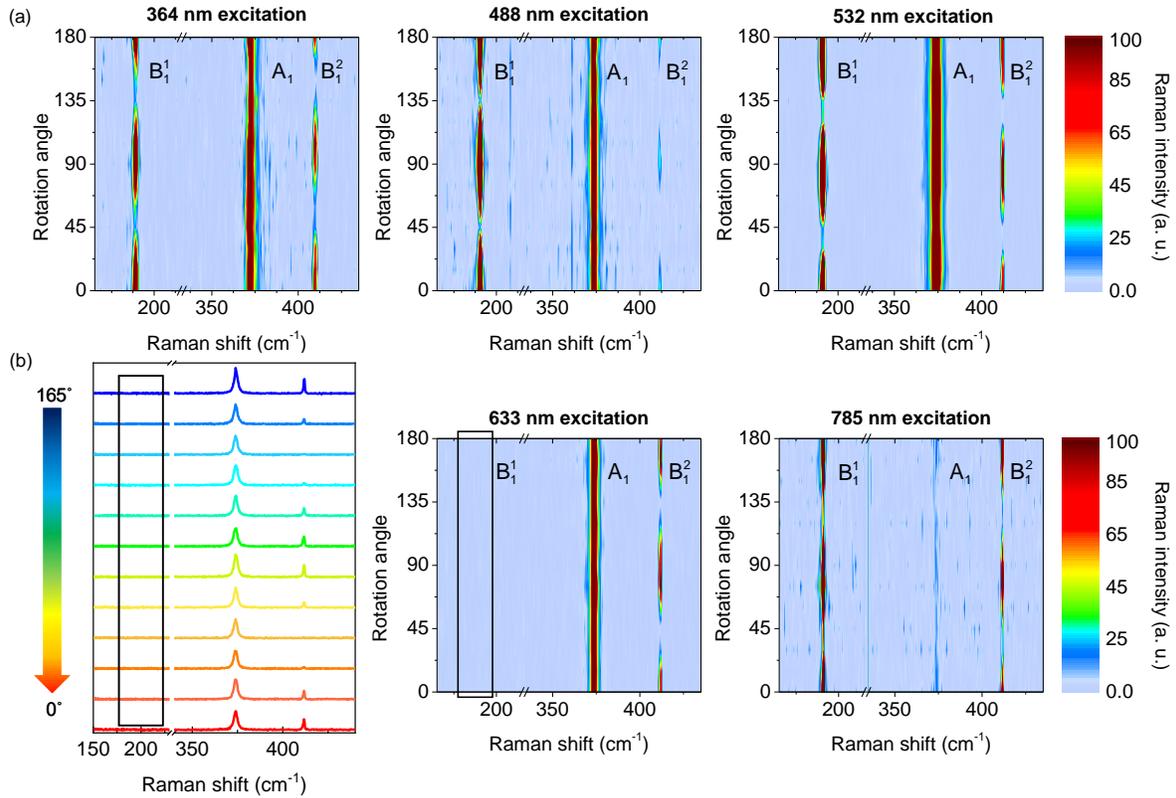

FIG. 3. Angle-resolved polarized Raman measurements on TaP. (a) Color image maps of the relative Raman intensity for 364, 488, 532, 633, and 785 nm excitations. The x-axes are the Raman shift and the y-axes are the polarization angles $\theta$ of the light with respect to the **a**-axis of TaP crystal from $\theta = 0°$ to $180°$. (b) Measured Raman spectra for 633 nm excitation from $\theta = 0°$ to $165°$. The $B_1^1$ mode at 190.5 cm$^{-1}$ is missing for 633 nm excitation at all the angles measured.

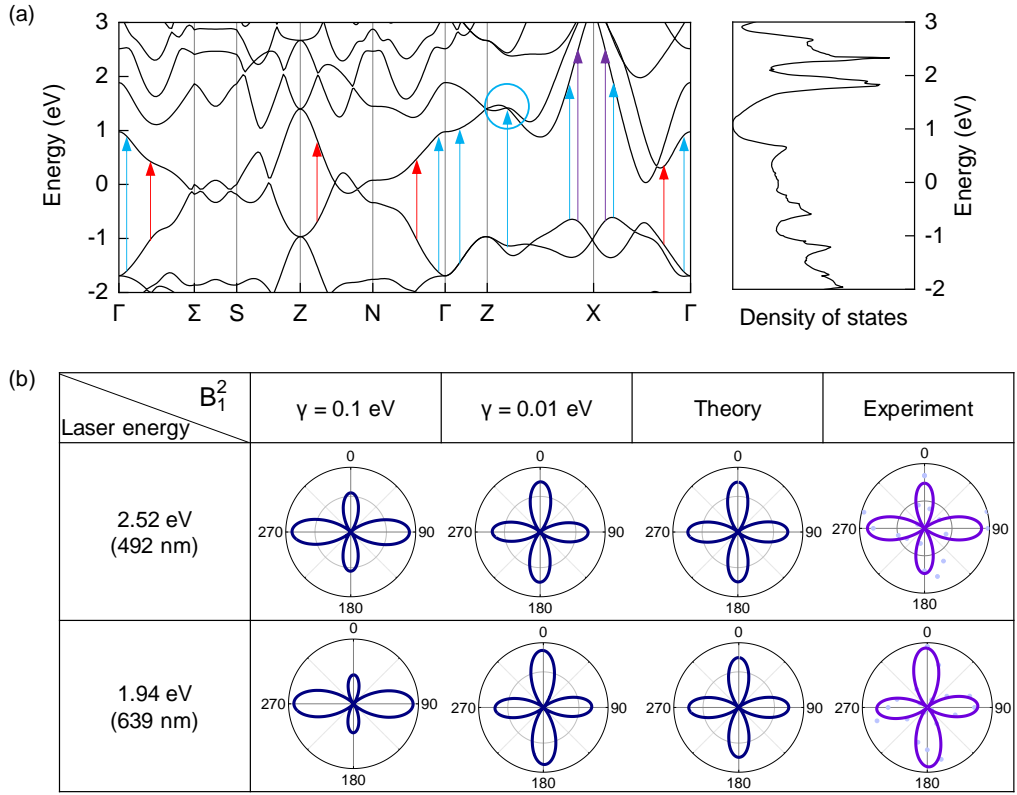

FIG. 4. $\gamma$ value dependence of TaP. (a) Excitations for 1.56 (red arrows), 2.52 (blue arrows), and 3.39 eV (purple arrows). For 2.52 eV, the conduction bands circled are degenerate. The right panel is the density of states. (b) Calculated polarization dependence for two different $\gamma$ values of 0.1 eV and 0.01 eV, classical Raman tensor theory and experiment of the polarization dependence for 2.52 and 1.94 eV laser excitation. The intensity scales of the polar plots are adjusted for clearer view.

TABLE I. Measured polar plots of the $B_1^1$, $A_1$ and $B_1^2$ modes for different laser excitations.

| Raman shift (cm⁻¹) | 190.5 | 373.9 | 411.7 |
|---|---|---|---|
| Symmetry | $B_1^1$ | $A_1$ | $B_1^2$ |
| 3.41 eV (364 nm) | 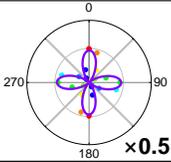 ×0.5 | 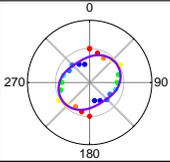 | 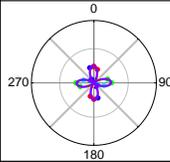 |
| 2.54 eV (488 nm) | 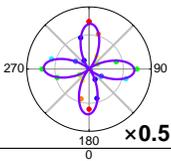 ×0.5 | 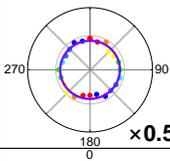 ×0.5 | 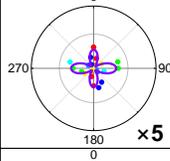 ×5 |
| 2.33 eV (532 nm) | 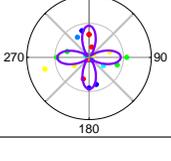 | 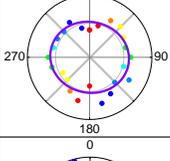 | 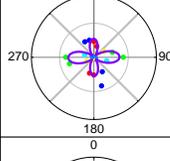 |
| 1.96 eV (633 nm) | No Peaks | 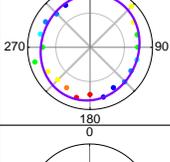 | 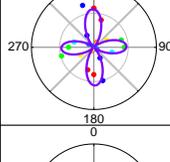 |
| 1.58 eV (785 nm) | 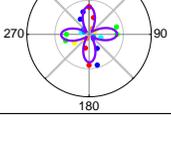 | 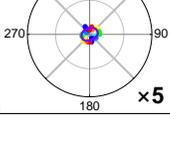 ×5 | 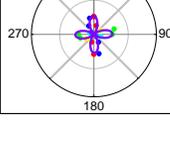 |

TABLE II. Calculated polar plots of the $B_1^1$, $A_1$ and $B_1^2$ modes for different laser excitations.

| Raman shift (cm⁻¹) | 179.5 | 355.0 | 395.2 |
|---|---|---|---|
| Symmetry | $B_1^1$ | $A_1$ | $B_1^2$ |
| 3.39 eV (366 nm) | 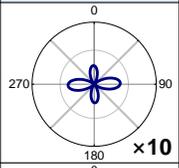 ×10 | 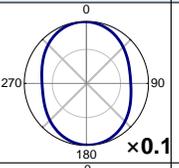 ×0.1 | 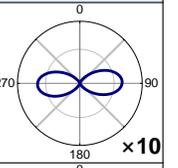 ×10 |
| 2.52 eV (492 nm) | 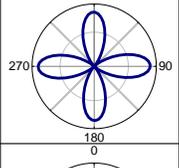 | 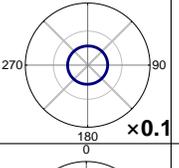 ×0.1 | 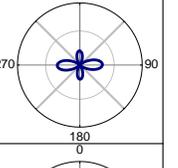 |
| 2.31 eV (536 nm) | 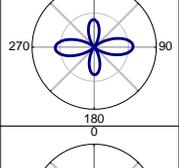 | 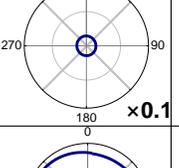 ×0.1 | 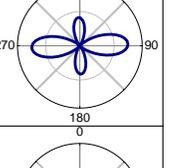 |
| 1.94 eV (639 nm) | 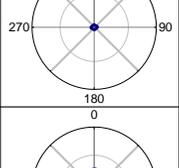 | 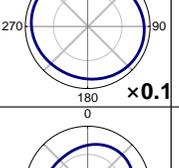 ×0.1 | 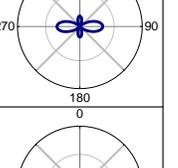 |
| 1.56 eV (794 nm) | 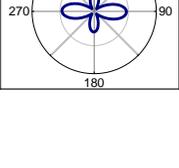 | 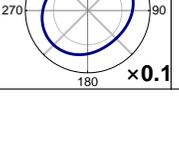 ×0.1 | 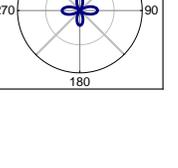 |

# Supplemental Material

# Anomalous Phonon-mode Dependence in Polarized Raman

# Spectroscopy of Topological Weyl Semimetal TaP

*Kunyan Zhang[†], Xiaoqi Pang[‡], Tong Wang[‡], Fei Han[§], Shun-Li Shang[∥], Nguyen T. Hung[⊥], Ahmad R. T. Nugraha[‡,#], Zi-Kui Liu[∥], Mingda Li[§,*], Riichiro Saito[‡,*] and Shengxi Huang[†,*]*

[†]Department of Electrical Engineering, The Pennsylvania State University, University Park, PA 16802, United States

[‡]Department of Physics, Tohoku University, Sendai 980-8578, Japan

[§]Department of Nuclear Science and Engineering, Massachusetts Institute of Technology, Cambridge, MA 02139, United States

[∥]Department of Materials Science and Engineering, The Pennsylvania State University, University Park, PA 16802, United States

[⊥]Frontier Research Institute for Interdisciplinary Sciences, Tohoku University, Sendai 980-8578, Japan

[#]Research Center for Physics, Indonesian Institute of Sciences, Tangerang Selatan 15314, Indonesia

1. X-ray diffraction of TaP.

It is important to note that the formation of twin crystal of TaP can be discarded as the cause of the disparity of Raman intensity along the **a-** and **b-** axes. In order to avoid the formation of twin crystal, we have optimized the growth method, which reduces the number of nucleation center significantly. Further, we intentionally choose the flat facet without stacking structure to perform the Raman measurement, since the type of twinning in TaP is penetration twinning and the twin boundary can be easily identified under optical microscopy.

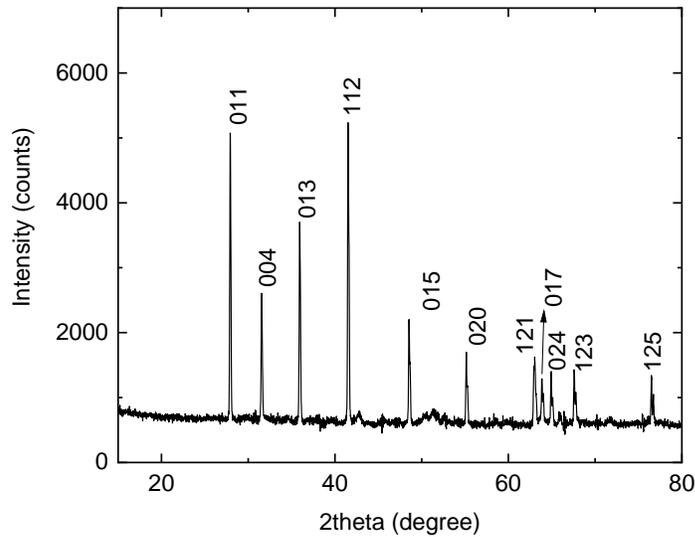

FIG. S1. X-ray diffraction of TaP.

2. Raman spectra and analysis of full width at half maximum (FWHM).

The FWHM of the peaks are analyzed here. There is no significant polarization dependence of FWHM for all the excitation wavelengths as shown in Fig. S3(b-f). Also, we need to note that the fitting error is large when the Raman intensity approximates to zero at certain angles (for example, at 45 degree for the $B_1^1$ mode), or the mode has inherently small Raman cross section (the $A_1$ mode for 785 nm excitation). FWHM of the Raman modes can be influenced by a series of physical parameters during measurements such as temperature and wavelength. Fig. S3(a) shows the measured excitation wavelength dependence of FWHM average by all the angle measured.

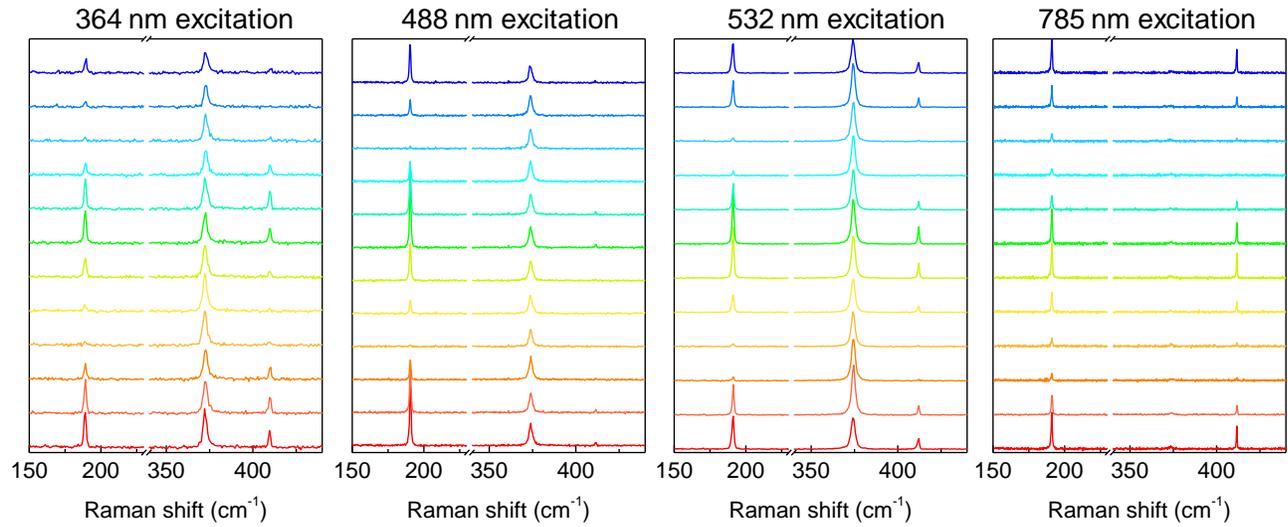

FIG. S2. Angle-resolved polarized Raman spectra for 364, 488, 532, 633 nm lasers for every 15˚ rotation angle.

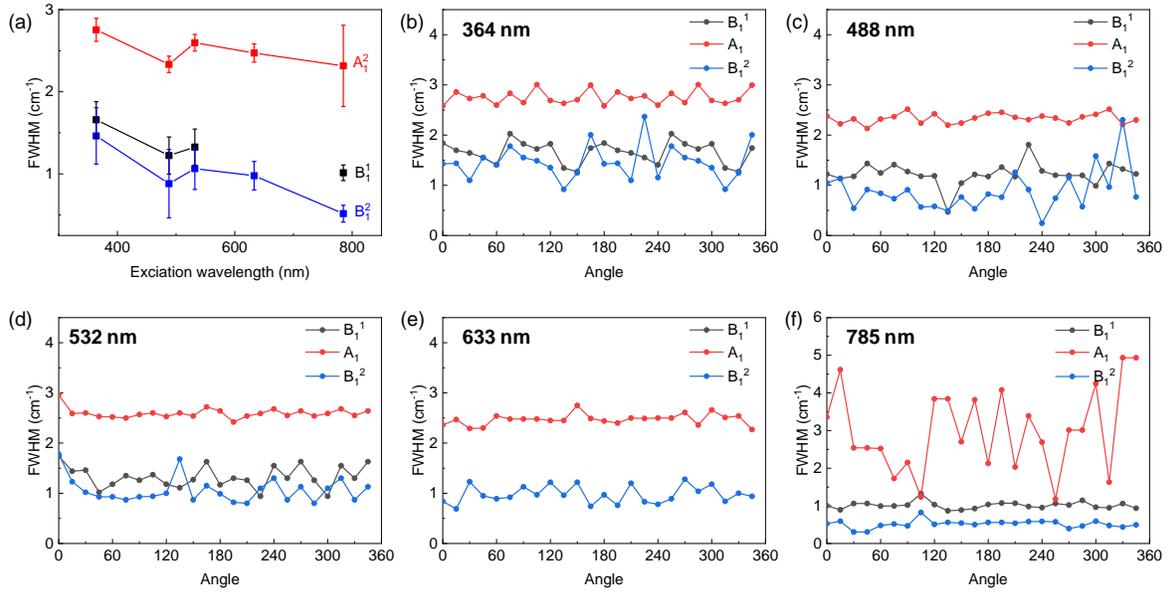

FIG. S3. FWHM of the Raman modes of TaP. (a) Excitation dependence of the FWHM. (b-f) Polarization dependence of the FWHM for excitation wavelength of 364, 488, 532, 633 and 785 nm.

3. Calculated Raman intensity by selecting the cases of $m' = m$ in the summation of $m'$ of Eq. (3).

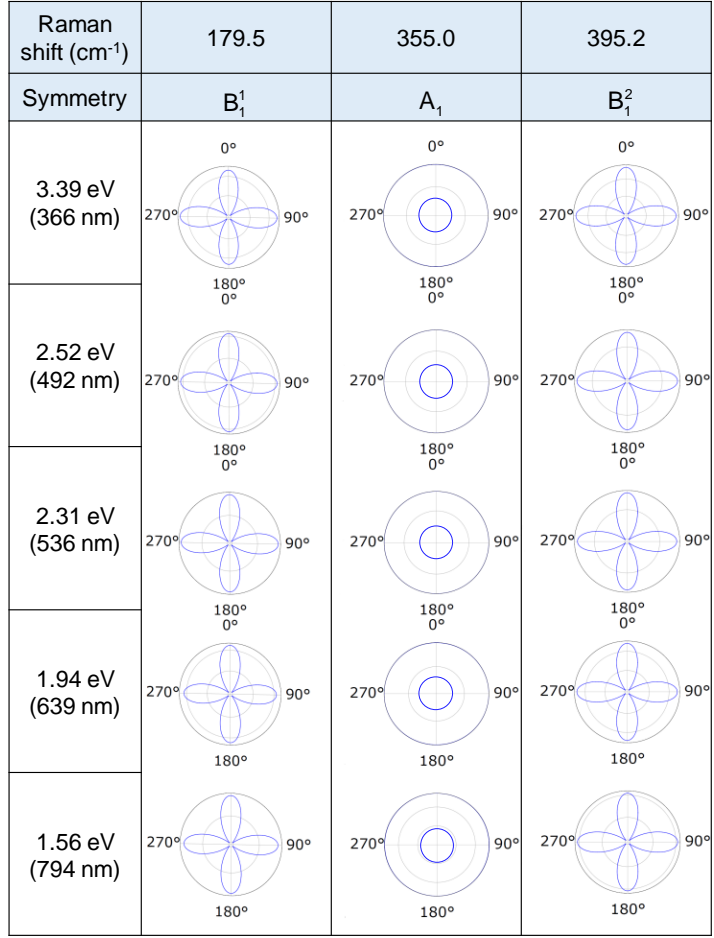

FIG. S4. The calculated polarized Raman spectra of the $B_1^2$ mode do not show any deviation from the symmetry of phonon when we intentionally select $(m, m') = $ (C1, C1), and (C2, C2) in Eq. (3) in the resonant Raman intensity calculation.

4. Calculated Raman intensity by Raman tensor theory.

TABLE SI. Raman tensor analysis of space group $I4_1md$. Six experimental configurations are considered with both parallel and perpendicular polarizations of the incoming and scattered light.

| Mode symmetry | A | $B_1^1$ | $B_1^2$ | E | |
|---|---|---|---|---|---|
| Raman tensor | $\begin{pmatrix} a & 0 & 0 \\ 0 & a & 0 \\ 0 & 0 & b \end{pmatrix}$ | $\begin{pmatrix} c & 0 & 0 \\ 0 & -c & 0 \\ 0 & 0 & 0 \end{pmatrix}$ | $\begin{pmatrix} 0 & d & 0 \\ d & 0 & 0 \\ 0 & 0 & 0 \end{pmatrix}$ | $\begin{pmatrix} 0 & 0 & e \\ 0 & 0 & 0 \\ e & 0 & 0 \end{pmatrix}$ | $\begin{pmatrix} 0 & 0 & 0 \\ 0 & 0 & e \\ 0 & e & 0 \end{pmatrix}$ |
| $Z(XX)\bar{Z}$ | $|a|^2$ | $|c\cos(2\theta)|^2$ | $|d\sin(2\theta)|^2$ | 0 | |
| $Z(XY)\bar{Z}$ | 0 | $|c\sin(2\theta)|^2$ | $|d\cos(2\theta)|^2$ | 0 | |
| $X(YY)\bar{X}$ | $|b\sin^2\theta + a\cos^2\theta|^2$ | $|c\cos^2\theta|^2$ | 0 | 0 | $|e\sin(2\theta)|^2$ |
| $X(YZ)\bar{X}$ | $|(a-b)\sin\theta\cos\theta|^2$ | $|c\sin\theta\cos\theta|^2$ | 0 | 0 | $|e\cos(2\theta)|^2$ |
| $Y(ZZ)\bar{Y}$ | $|a\sin^2\theta + b\cos^2\theta|^2$ | $|c\sin^2\theta|^2$ | 0 | $|e\sin(2\theta)|^2$ | 0 |
| $Y(ZX)\bar{Y}$ | $|(a-b)\sin\theta\cos\theta|^2$ | $|c\sin\theta\cos\theta|^2$ | 0 | $|e\cos(2\theta)|^2$ | 0 |